**Hemoglobin Non-equilibrium Oxygen Dissociation Curve.**

Rosella Scrima[1], Sabino Fugetto[1], Nazzareno Capitanio[1], Domenico L. Gatti[2]

[1] Department of Clinical and Experimental Medicine, University of Foggia, Via L. Pinto 1, 71122 Foggia, Italy.
[2] Department of Biochemistry, Microbiology and Immunology, Wayne State University School of Medicine, 540 E. Canfield Avenue, Detroit, MI, USA.

Corresponding author: Domenico Gatti, dgatti@med.wayne.edu

**Abstract:** Abnormal hemoglobins can have major consequences for tissue delivery of oxygen. Correct diagnosis of hemoglobinopathies with altered oxygen affinity requires a determination of hemoglobin oxygen dissociation curve (ODC), which relates the hemoglobin oxygen saturation to the partial pressure of oxygen in the blood. Determination of the ODC of human hemoglobin is typically carried out under conditions in which hemoglobin is in equilibrium with $O_2$ at each partial pressure. However, in the human body due to the fast transit of RBCs through tissues hemoglobin oxygen exchanges occur under non-equilibrium conditions. We describe the determination of non-equilibrium ODC, and show that under these conditions Hb cooperativity has two apparent components in the Adair, Perutz, and MWC models of Hb. The first component, which we call *sequential cooperativity*, accounts for ~70% of Hb cooperativity, and emerges from the constraint of sequential binding that is shared by the three models. The second component, which we call *conformational cooperativity*, accounts for ~30% of Hb cooperativity, and is due either to a conformational equilibrium between low affinity and high affinity tetramers (as in the MWC model), or to a conformational change from low to high affinity once two of the tetramer sites are occupied (Perutz model).

**1. Introduction.**
Hemoglobin (Hb) oxygen dissociation curve (ODC), which relates oxygen saturation ($S_{O2}$) and partial pressure of oxygen in the blood ($P_{O2}$), is an important tool for understanding how blood carries and releases oxygen [1].

Classically, factors recognized to influence the ODC include the local $CO_2$ partial pressure ($P_{CO2}$), pH, temperature, as well as allosteric metabolites like 2,3 diphosphoglycerate (2,3-DPG). The curve is shifted to the right (i.e. lower saturation for a given $P_{O2}$) by higher $P_{CO2}$, greater acidity (lower pH), higher temperature, and higher concentration of 2,3-DPG [1-7]. The factors that shift the ODC to the right are directly relevant to the conditions which prevail in metabolizing tissues, as they facilitate the unloading of oxygen from hemoglobin. The converse occurs during passage through the pulmonary capillaries, with the greater affinity accompanying a shift of the ODC to the left aiding the uptake of oxygen [8].

The partial pressure of oxygen in the blood at which the hemoglobin is 50% saturated is known as the $P_{50}$. The $P_{50}$ of normal hemoglobin is approximately 26 mmHg at a partial $CO_2$ pressure of 40 mmHg [7]. In the presence of disease or other conditions that change hemoglobin oxygen affinity and, consequently, shift the curve to the right or left, the $P_{50}$ changes accordingly [9]. Low affinity hemoglobins are characterized by higher $P_{50}$ and high-affinity hemoglobins by a lower $P_{50}$.

Hemoglobin ODC was originally determined by manual methods in which the oxygen saturation of hemoglobin was measured by spectrophotometry at every stepwise change of $P_{O2}$ by the addition of aliquots of air to the pre-deoxygenated sample in a tonometer, after equilibrium between Hb and $O_2$ at each $P_{O2}$ was reached. Although very accurate, this *static* method was very laborious and time consuming. To overcome its limitations, *dynamic* methods were developed in which $P_{O2}$ was progressively changed in a close vessel, allowing for equilibration in between changes [1,10-14]. In the most current implementation of these methods, a Clark oxygen electrode detects the change in oxygen tension, while the resulting increase in oxyhemoglobin fraction is simultaneously monitored by dual-wavelength spectrophotometry at 560 nm and 576 nm [15].

However, equilibrium ODC does not reflect exactly physiological gas exchanges by Hb in the lungs (oxygenation) and in peripheral tissues (deoxygenation) as, due to the fast transit of RBCs through microcirculation [16], hemoglobin $O_2$ exchanges occur under non-equilibrium conditions. We describe the determination and features of non-equilibrium ODC and the contribution of different components of Hb cooperativity to its shape.

**2. Material and Method.**
Rat liver mitochondria were purified by differential centrifugation of tissue homogenate as described in [17], portioned in aliquots of 30-40 mg protein/ml in 0.25 mM sucrose, and stored at -80 °C till used. *Stripped* Hb devoid of allosteric heterotropic factors was prepared from whole human blood of healthy donors as in [18] and its concentration, as $HbO_2$, was estimated spectrophotometrically from its heme content, with an $\varepsilon_{mM}$ at 577 nm of 15.4 [19]. Alternatively, Vacutainer (Becton, Dickinson and company) collected whole blood was used; in this case the $HbO_2$ concentration was estimated, after dilution of the sample in double distilled water, from the air-equilibrated minus $Na_2S_2O_4$-supplemented differential spectra, using a $\Delta\varepsilon_{mM}$ at 577-568 nm of 4.8 according to [20]; the Hb content (as heme centers) was typically 9-11 nmol/(ml blood).

Determination of oxygen consumption was carried out by respirometry using Oxygraph-2k (O2k, OROBOROS Instruments, Innsbruck, Austria). The instrument has two measuring chambers (2 ml each) both equipped with a Clark-type electrode; calibration of the instrument was performed according to the manufacturer instructions and all the measurements were carried out at 37 °C. The assay medium constituted by 250 mM sucrose, 1 mg/ml bovine serum albumin, 10 mM $KH_2(PO_4)$, 27 mM KCl, 1 mM $MgCl_2$, 40 mM Hepes, 0.5 mM EGTA (pH 7.4) was supplemented with 20-50 µg/ml mitochondrial proteins in the absence or presence of either purified Hb or whole blood samples. Oxygen consumption was initiated by the addition of 10 mM succinate as respiratory substrate in the presence of 2 µM rotenone, an inhibitor of the respiratory chain NADH-ubiquinone oxidoreductase/Complex I. Alternatively the measurements were carried out with a single-chamber (0.5-1 ml) oxymeter equipped with a Clark type electrode disc (Hansatech Instruments Ltd, King's Lynn, UK).

Fitting of the deoxygenation/reoxygenation curves of Hb using either a *graphic* (see Supplementary Information) or *kinetic* method (Results section) was carried out using in-house code written for Matlab® (deposited at https://github.com/dgattiwsu/HB_ODC). Kinetic models of Hb (Adair, MWC, Perutz) were built using functions of Matlab basic functionalities and of the Optimization Toolbox. Using these models, a single set of rate constants was derived by global analysis of both the deoxygenation and reoxygenation traces.

## 3. Results.

**3.1. Non-equilibrium ODC of human hemoglobin.** In the following we describe a typical experimental determination of human hemoglobin non-equilibrium ODC.

2 ml of the assay buffer solution (pH 7.4) were placed inside each of the two glass chambers of the Oroboros oxymeter and supplemented with a small amount of purified rat liver mitochondria to a final concentration of 30 μg prot/ml. With the cell open to the environment, the solutions were allowed to equilibrate under stirring with atmospheric $O_2$ until the observed $O_2$ concentration (based on an earlier calibration of the electrode [21-23]) remained stable (~183 μM) for a few minutes.

Upon insulating the glass cells from air with glass stoppers (containing a small port for microsyringe additions) progressive reduction of the $P_{O2}$ in the cell was achieved by adding 10 mM succinate according to the reaction catalyzed by the succinate oxidase segment of the respiratory chain (complexes II+III+IV: succinate dehydrogenase, ubiquinol:cytochrome c reductase, cytochrome c oxidase, respectively):

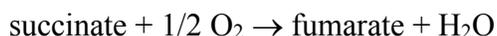

$$\text{succinate} + 1/2\ O_2 \rightarrow \text{fumarate} + H_2O$$

The final concentration of *cytochrome c oxidase* (the mitochondrial enzyme that uses up $O_2$ reducing it to water) in the cell was estimated to be ~0.05 μM by visible spectroscopy of the mitochondrial suspension in the 500-650 nanometer range. Upon starting respiration, $O_2$ concentration decreased linearly until anaerobiosis was reached (**Fig. 1a**).

The linear respiratory activity is due to the affinity of cytochrome *c* oxidase for $O_2$ whose $K_M$ is estimated in the submicromolar range [24,25]. Considering that the instrumental limit of detection of $O_2$ concentration is also in the submicromolar range [26], the $O_2$ concentration is practically never limiting the respiratory flux under the prevailing conditions reported here. After adding 5 μM Antimycin A, an inhibitor of respiration at the level of *ubiquinol:cytochrome c reductase*, the glass stopper was removed, and the cell content was allowed to equilibrate again with air. During this phase oxygen diffuses back in the cell in a non-linear fashion.

When the cycle of deoxygenation/reoxygenation is repeated in the presence of 50 μM (as heme centers) human hemoglobin isolated from blood hydrolysate (**Fig. 1b**) the final part of the deoxygenation curve is no longer linear due to the release of oxygen from Hb (green arrow). The initial part of the reoxygenation curve is also slower due to the uptake of oxygen by Hb (yellow arrow).

**Fig. 1**. **a.** Polarographic trace of deoxygenation triggered by succinate, and reoxygenation triggered by reopening the Clark cell in the presence of Antimycin A. **b.** Deoxygenation / reoxygenation cycle in the presence of 50 µM Hb. A green and a yellow arrow point to the deviations from the pattern seen in **a**.

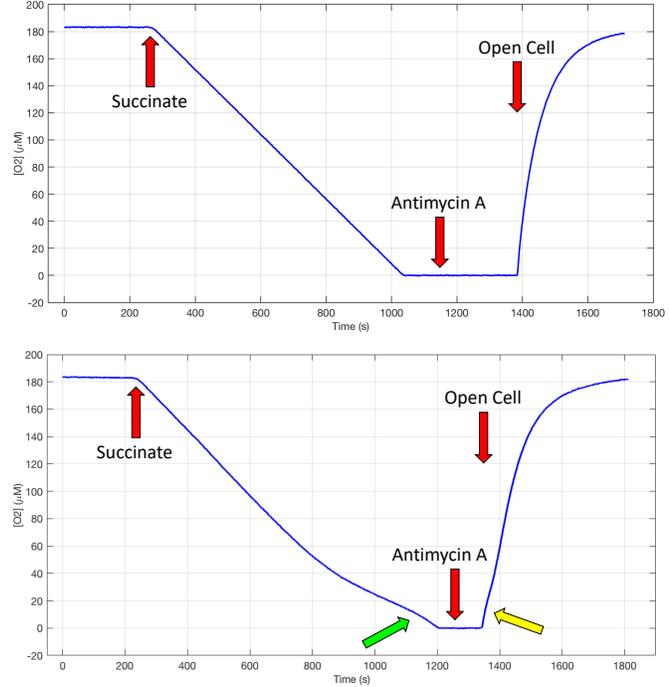

Since the alterations of the deox-/reoxygenation curves are due to the release/uptake of $O_2$ by Hb, it is possible to recover the non-equilibrium ODC using a kinetic model of the cell ensemble as a set of reversible reactions involving the species of Hb in different oxygenation state, *cytochrome c oxidase* as the terminal $O_2$ acceptor, $O_2$ in the cell, and external air in diffusive equilibrium with the $O_2$ in the cell, when the latter is open. We have evaluated three different kinetic models (**Fig. 2**): **a**. a <u>sequential</u> Adair style model with 4 refined parameters (one $O_2$ $k_{off}$ for each $Hb(O_2)_n$), **b**. a *two state sequential* Perutz style model with 2 refined parameters (one $O_2$ $k_{off}$ for $Hb(O_2)_{1,2}$ and one $O_2$ $k_{off}$ for $Hb(O_2)_{3,4}$), and **c**. a *two state concerted* Monod-Wyman-Changeux (MWC) style model (with one $K_{equil}$ between a <u>tense</u> (T) and a <u>relaxed</u> (R) state of Hb, one $O_2$ $k_{off}$ for all T states and one $O_2$ $k_{off}$ for all R states) (reviewed in [27,28]).

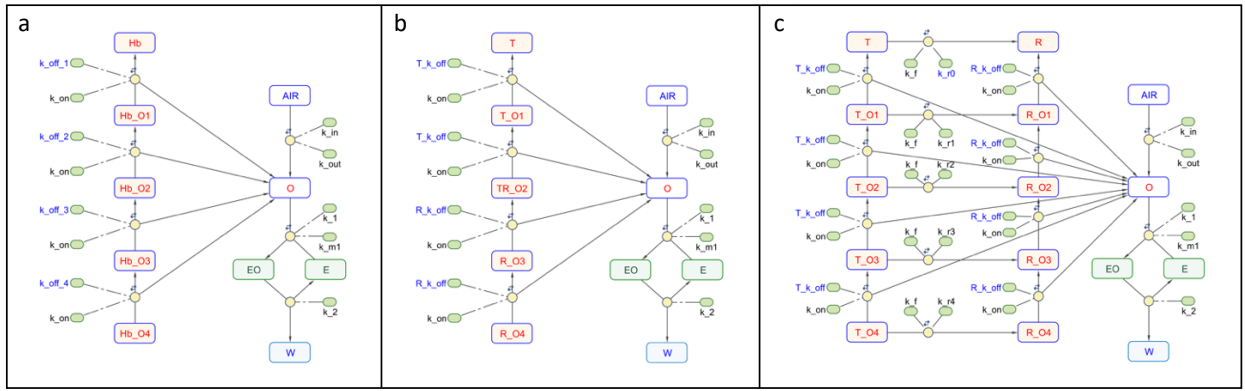

**Fig. 2.** Kinetic model of Hb ODC. **a.** Adair model. **b.** Perutz model. **c.** MWC model. Species are shown as rounded rectangles: E = cytochrome c oxidase; O = cell $O_2$; AIR = air $O_2$; EO = *cytochrome c oxidase*:$O_2$; W = water. Reactions are shown as yellow circles, rate constants as small rounded rectangles. Rate constants that are refined in the model are highlighted in blue font. A double arrow over each reaction circle indicates full reversibility. Initial values for all the *on* rate constants were set at 100 µM$^{-1}$s$^{-1}$; initial values for the *off* rate constants were set accordingly based on the values of the equilibrium constants for the progressive binding of $O_2$ to Hb as derived from an initial fit of the oximetric traces with a simple graphic method (Supplementary Information) and using Adair equation [1,29,30]. Initial value for the $k_{cat}$ of *cytochrome c oxidase* were derived from the linear phase of respiration supported by succinate. In the Perutz model the species TR_O2 is used to represent the fast conformational transition from T_O2 to R_O2.

The three models fit equally well the oxymetric trace with essentially identical *sum of square errors* (~150 μM$^2$) and *R-square* values (>.99). Each model provides the contribution of all Hb species (**Fig. 3a,d,g**) at all time points (O2 concentrations). Non-equilibrium ODC's are derived for each P$_{O2}$ value as the ratio between the sum of all the oxygenated species and the total amount of Hb (**Fig. 3b,e,h**). The corresponding Hill plots are shown in **Fig. 3c,f,i**.

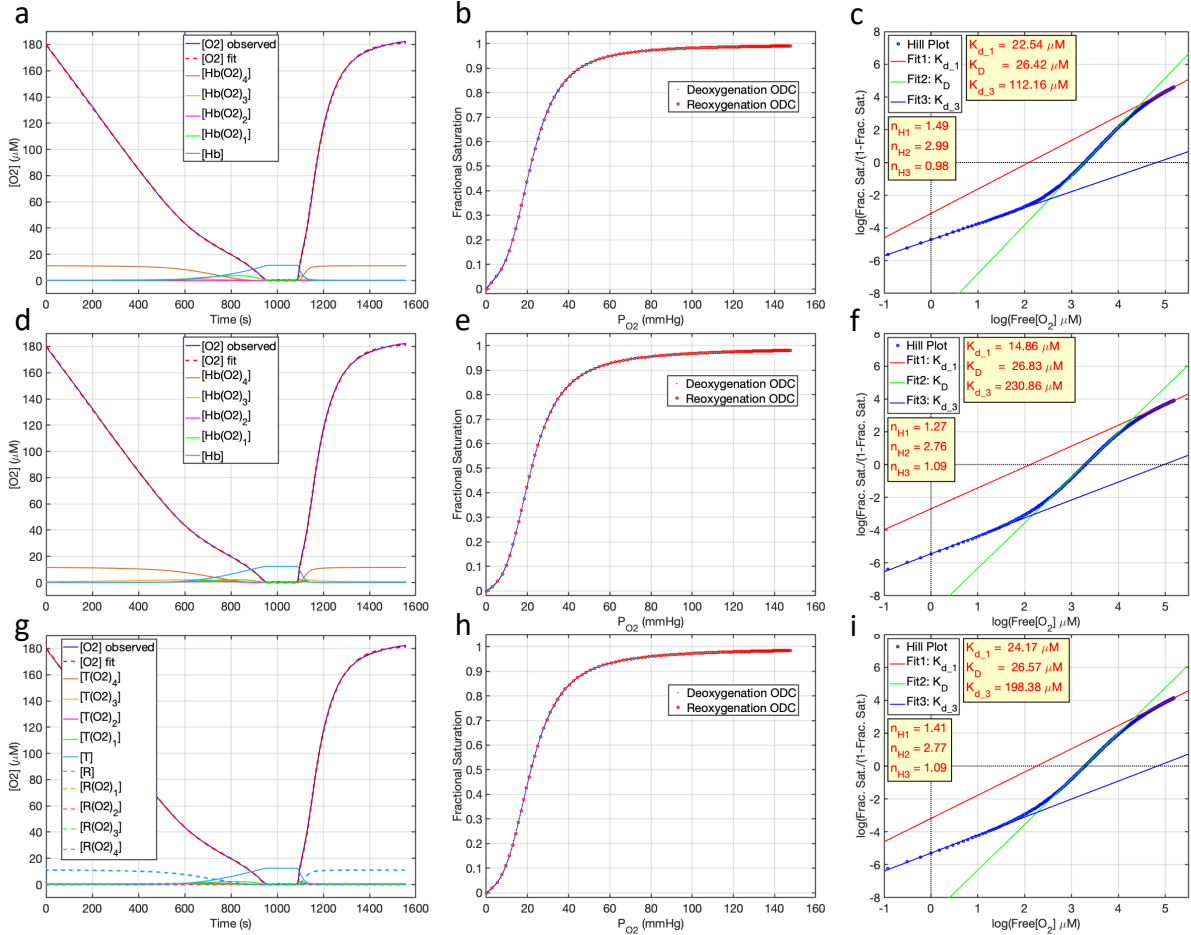

**Fig. 3. a,d,g.** Kinetic fit of the deoxygenation/reoxygenation cycle shown in Fig. **1b** using the Adair, Perutz, and MWC model, respectively. The refined concentration of Hb tetramers was 11.7, 12.4 , 12.5 μM in the Adair, Perutz, and MCW models, respectively. **b,e,h.** ODC's calculated from the fits in **a,d,g**, respectively, using only the deoxygenation (blue dots) or the reoxygenation (cyan circles with red outline) phase. The concentrations of the different Hb species refer to the entire tetramer: therefore, fractional saturation is calculated as ([Hb(O$_2$)$_4$] x 4 + [Hb(O$_2$)$_3$] x 3 + [Hb(O$_2$)$_2$] x 2 + [Hb(O$_2$)])/([Hb$_{total}$] x 4). The calculated P$_{50}$ from a non-linear least-squares fit of all the points in the ODC's with Adair equation is ~21.9 mmHg. **c,f,i.** Corresponding Hill plots calculated combining points from the deoxygenation and reoxygenation phase.

The observed variations in the ODC and Hill plots are due to the fact that for each time point the contributions of individual Hb species are different in the three models. Details of these contributions are shown in a blow-up of the terminal part of Hb deoxygenation phase (**Fig. 4a,c,e**), and of the initial part of Hb reoxygenation phase (**Fig. 4b,d,f**). Regardless of the model used, concentration peaks are reached in the order Hb(O$_2$)$_4$ → Hb(O$_2$)$_3$ → Hb(O$_2$)$_2$ → Hb(O$_2$)$_1$ → Hb during deoxygenation, and Hb → Hb(O$_2$)$_1$ → Hb(O$_2$)$_2$ → Hb(O$_2$)$_3$ → Hb(O$_2$)$_4$ during reoxygenation.

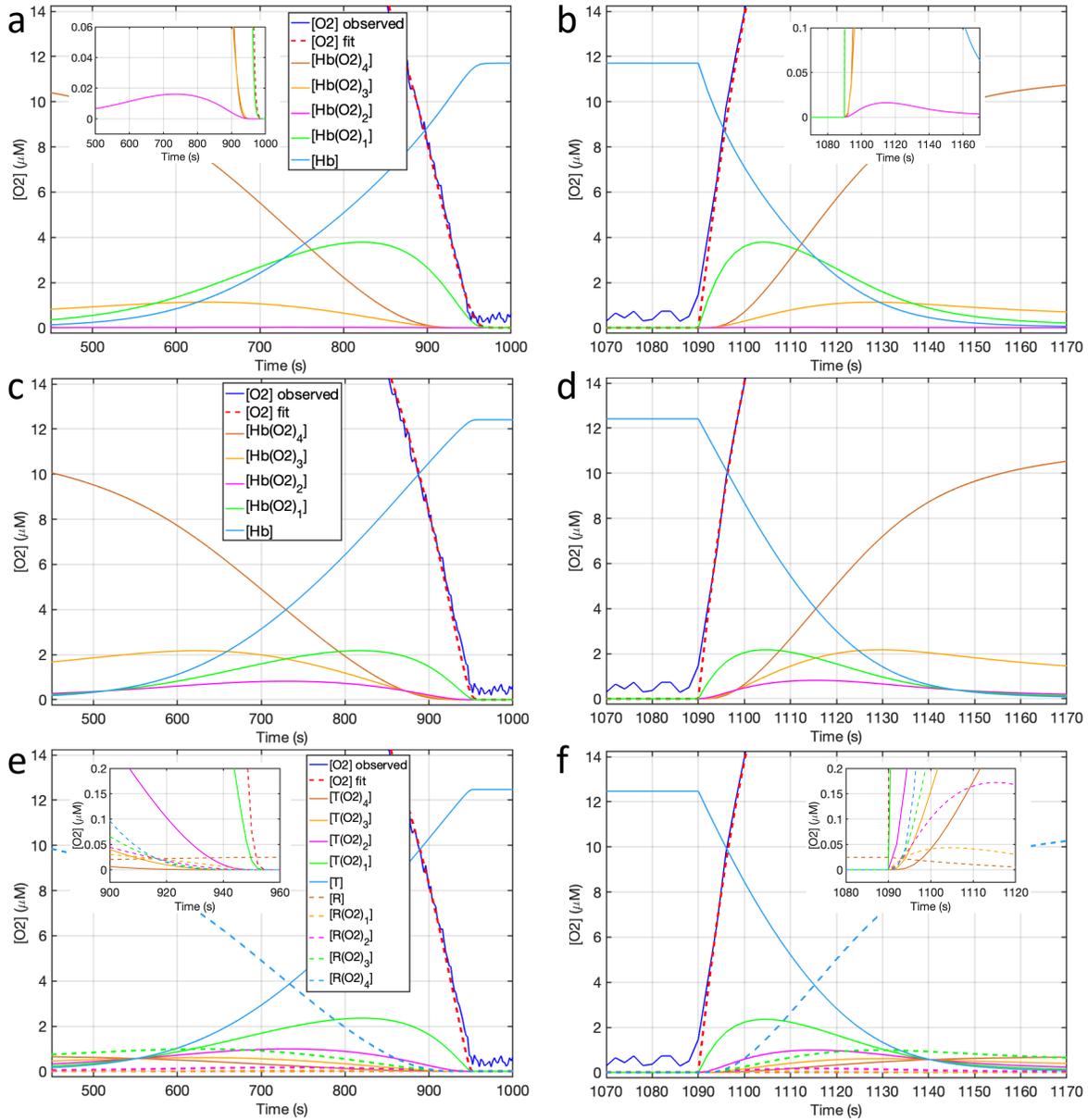

**Fig. 4.** Kinetic fit of the deoxygenation/reoxygenation cycle shown in Fig. **1b** using the Adair, Perutz, and MWC model, respectively. **a,c,e.** Blow-up of the terminal part of Hb deoxygenation phase. **b,d,f.** Blow-up of the initial part of Hb reoxygenation phase. Insets in **a,b** show the peaks of Hb(O$_2$)$_2$. Notice the cross-over points at 520 s (**c**) and 1146 s (**d**) between Hb, Hb(O$_2$)$_1$, Hb(O$_2$)$_2$, corresponding to Perutz T conformation, and at 870 s (**c**) and 1098 s (**d**) between Hb(O$_2$)$_2$, Hb(O$_2$)$_3$, and Hb(O$_2$)$_4$, corresponding to Perutz R conformation. Likewise, cross-over points are present at 917 s (**e, inset**) and 1093 s (**f, inset**) between all R states, and at 551 s (**e**) and 1040 s (**f**) between all T states of the MWC model.

In the Perutz and MWC models, intercepts with the X axis of the extrapolated lines from the asymptotic ends of the Hill plot are usually interpreted as the concentration of O2 at which the concentrations of the unliganded and liganded forms of the T and R states are equal ($K_D$). Better values for these magnitudes are calculated directly from the cross-over points (**Fig. 4**) of the T and R forms derived from the kinetic simulations (**Table 1**). Non-linear least-squares fit of Adair equation [30],

$$S_{O2} = \frac{K_1[O_2] + 2K_1K_2[O_2]^2 + 3K_1K_2K_3[O_2]^3 + 4K_1K_2K_3K_4[O_2]^4}{4(1 + K_1[O_2] + 2K_1K_2[O_2]^2 + 3K_1K_2K_3[O_2]^3 + 4K_1K_2K_3K_4[O_2]^4)}$$

$$K_1 = \frac{[Hb(O_2)]}{[Hb][O_2]} \; ; \; K_2 = \frac{[Hb(O_2)_2]}{[Hb(O_2)][O_2]} \; ; \; K_3 = \frac{[Hb(O_2)_3]}{[Hb(O_2)_2][O_2]} \; ; \; K_4 = \frac{[Hb(O_2)_4]}{[Hb(O_2)_3][O_2]},$$

to the combined experimental points from the deoxygenation and reoxygenation *non-equilibrium* ODC's can be used to derive values for the *equilibrium* Adair constants in the three models (**Table 1**).

| Table 1 | Imai *et al.*, 1994 [6] | Minimalist model | Adair model | Perutz model | MWC model |
|---|---|---|---|---|---|
| | | | | | c = 0.1069, $L_0$=504.6, $L_1$=53.9, $L_2$= 5.76, $L_3$= 0.62, $L_4$= 0.066, $K_T$ = 0.0191, $K_R$ = 0.1787 |
| $K_{D\_crossover}$ | | 27.95 (=1/$K_1$) | | | |
| $K_{D\_Hill}$ | | 27.73 | 26.42 | 26.83 | 26.57 |
| $K_DT_{\_crossover}$ | | | | 59.68 (=1/$K_1$) | 52.09 (=1/$K_T$) |
| $K_DR_{\_crossover}$ | | | | 12.04 (=1/$K_4$) | 5.69 (=1/$K_R$) |
| $K_1$ | 0.0188 | 0.0360 | 0.0368 | 0.0168 | 0.0194 |
| $K_2$ | 0.0566 | 0.0360 | 0.0002 | 0.0168 | 0.0220 |
| $K_3$ | 0.4070 | 0.0360 | 2.2237 | 0.0818 | 0.0428 |
| $K_4$ | 4.2800 | 0.0360 | 0.1730 | 0.0818 | 0.1178 |

**Table 1.** $K_D$'s (µM) from model refinement, and Adair equilibrium (association) constants (µM$^{-1}$) for the four steps of Hb oxygenation from a fit of Adair equation to the *non-equilibrium* ODC's. For the Adair and Perutz model, these values are identical to those refined during the optimization of the two models. Notice the collapse of 4 Adair constants into just 2 in the Perutz model, and just 1 in the Minimalist model (see below, Sources of cooperativity in the models). In the MWC model, $K_T$ and $K_R$ are the refined association constants for the T and R states oxygenation, c = $K_T/K_R$, and $L_0$ = [$T_0$]/[$R_0$] is the refined allosteric constant expressing the equilibrium in the absence of any oxygenation between the T and R state of Hb. Allosteric constants $L_n$ for the *n* different oxygenation levels are not refined, but calculated according to the relationship $L_n = L_0 c^n$.

## 4. Sources of cooperativity in the models.

A common feature of Adair, Perutz, and MWC models is the presence of sequential reactions. A *minimalist model* containing four sequential binding reactions, $Hb \leftrightarrow Hb(O_2)_1 \leftrightarrow Hb(O_2)_2 \leftrightarrow Hb(O_2)_3 \leftrightarrow Hb(O_2)_4$, with a single $O_2$ affinity for all states, and no conformational changes (**Fig. 5a**), fits experimental observations surprisingly well ($sse$ = 515.5, $R^2$ = 0.998) (**Fig. 5b,c,d**), giving origin to a sigmoidal ODC (**Fig. 5e**) and a Hill plot with asymptotic components that suggest the presence of both low and high affinity sites (**Fig. 5c**), despite none such exist in the model. Accordingly, a fit of Adair equation to the sigmoidal ODC shows that all four Adair constants collapse into a single one corresponding to the refined $k_{on}/k_{off}$ ratio (**Table 1**). This observation suggests that a first source of cooperativity in the Adair, Perutz, and MWC models might reside in the fact that all three models feature *sequential* binding reaction of $O_2$ to the four sites of Hb. A single additional conformational change between a *low* and *high* affinity state, whether it occurs upon occupancy of the first two sites (Perutz), or at every level of $O_2$ occupancy, provides the remaining character of cooperativity to $O_2$ binding.

We define as the *cooperativity gain* (*cgain*) of a model the *rms* (root mean square) deviation between the model derived ODC and the ODC with the same $P_{50}$ derived from a model containing 4 independent identical $O_2$ sites. The *cooperative gain* of the minimalist, Perutz and NWC models is shown in **Fig. 5e,g,h**.

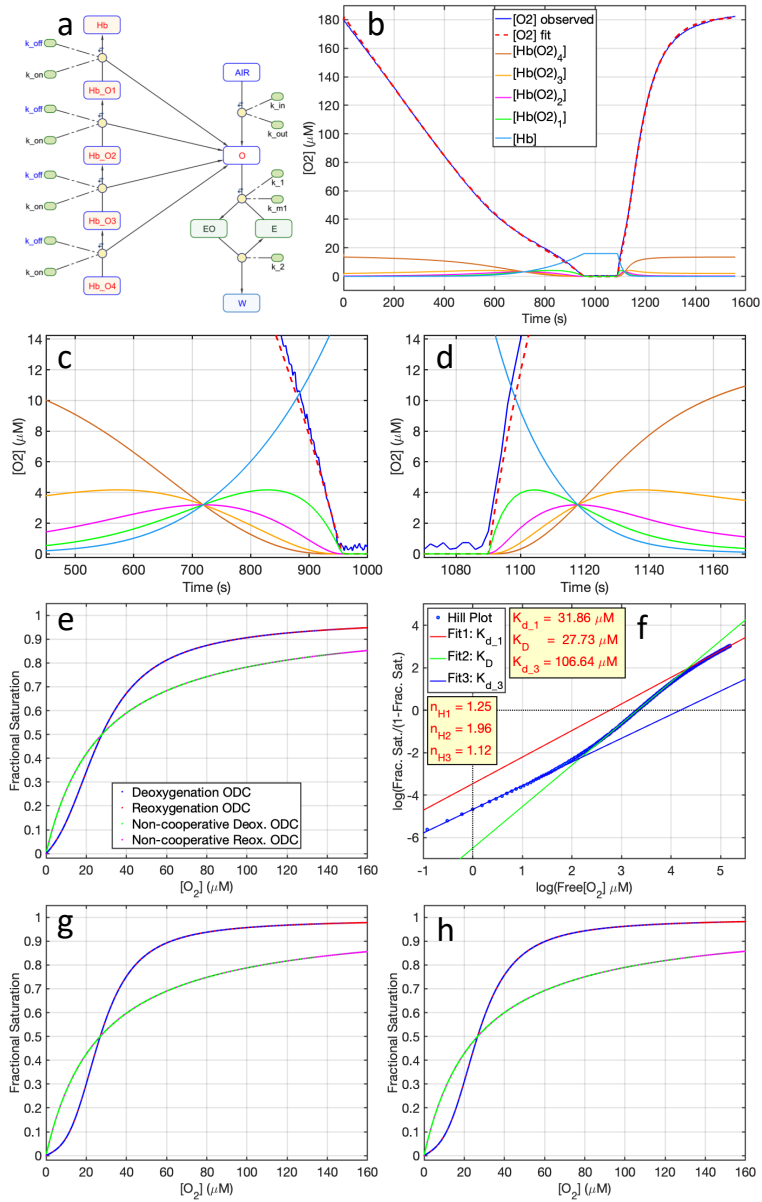

**Fig. 5. a,** Minimalist model with the same $O_2$ affinity at all binding sites. **b**, Kinetic fit of the deox/reox cycle shown in Fig. **1b**. **c,d.** Blow-up of the terminal part of Hb deoxygenation phase and the initial part of Hb reoxygenation phase. **e,** ODC calculated from the fit in panel **b**, using only the deoxygenation (blue dots) or the reoxygenation (red circles) phase. The ODC with the same $P_{50}$ derived from a model containing 4 independent identical $O_2$ sites is shown for the same phases as green and magenta dots (*cgain* = 0.1052). **f**, Corresponding Hill plot calculated combining points from the deoxygenation and reoxygenation phase. **g,h,** ODC's showing the *cooperativity gain* of the Perutz (*cgain* = 0.1505) and MWC models (*cgain* = 0.1526).

## 5. Discussion.

We have presented a typical experiment showing the determination of hemoglobin non-equilibrium ODC. The experimental component of the method requires the acquisition of a deoxygenation curve of Hb, and optionally, also of a reoxygenation curve. The computational component is based on the minimization of the *sum of square errors* (*sse*) between an experimentally observed $O_2$ polarographic trace and a simulated $O_2$ trace based on a kinetic model of choice (**Figs. 2-4**). Since it does not require an optical determination of the Hb saturation, this method can be directly used with a red cell suspension or whole blood without the added complications of the dual-wavelength or full sphere spectrometry that are necessary to eliminate light scattering noise [10].

It has not escaped our notice that the equilibrium constants calculated with our method for each Hb oxygenation step are significantly different from those reported by other authors (i.e. [1], **Table 1**). It is important to recognize that while traditional ODC's are derived under conditions of *full equilibrium* between Hb and $O_2$ at different values of $P_{O2}$, our method, based on a continuous change of the $O_2$ concentration due to mitochondrial respiration or $O_2$ diffusion from the air, derives a *non-equilibrium* ODC, that possibly reflects more closely the physiological condition of red blood cells moving rapidly in the blood stream across regions of different $P_{O2}$.

Three kinetic models (Adair *sequential*, Perutz *two state sequential*, MWC *two state concerted*) were all equally effective in fitting the experimental polarographic traces. Thus, as such, the methodology presented here does not offer any new ways to discriminate between these models. However, the predictions made by the three models with respect to the concentrations of oxygenation and conformational intermediates are quite different, and may offer inspiration for future experiments.

The origin of cooperativity in Hb has been a point of intense debate for over a century (reviewed in [27,28,31]). Our experiments were carried out with *stripped* Hb devoid of allosteric heterotropic factors, and thus under these conditions we observed the intrinsic cooperative behavior of tetrameric Hb. A *minimalist* model, containing no induced (by $O_2$ binding) or intrinsic conformational equilibria between a *low* and a *high* affinity state, is sufficient to provide, by virtue of the constraint of *sequential* binding reactions of identical affinity, a large fraction (~69% in *cgain* scale) of the cooperative behavior of Hb, as judged by the magnitude of its *cooperative gain* (= 0.1052) with respect to a model containing 4 independent identical $O_2$ binding sites. We call this component the **sequential cooperativity** of Hb. A smaller component (~31% in *cgain* scale) of cooperativity is provided by inclusion in the model of a single conformational equilibrium between a *low* and a *high* affinity state, as shown by the values of the *cooperative gain* in the Perutz (*cgain* = 0.1505) and MWC models (*cgain* = 0.1526). We call this component the **conformational cooperativity** of Hb. Together, *sequential* and *conformational* cooperativity account for the entire *cooperativity gain* of Hb under non-equilibrium conditions.

**Literature cited.**

**Declarations of interest**. None.

**Data repository**. Test data and Matlab scripts to carry out the graphic and/or kinetic method for ODC determination are freely available at https://github.com/dgattiwsu/HB_ODC.

**Funding**. This work was supported by a Wayne State University Research Enhancement Program in Computational Biology grant to DLG, and by grants from the University of Foggia to RS, and NC.

**Supplementary Information.**

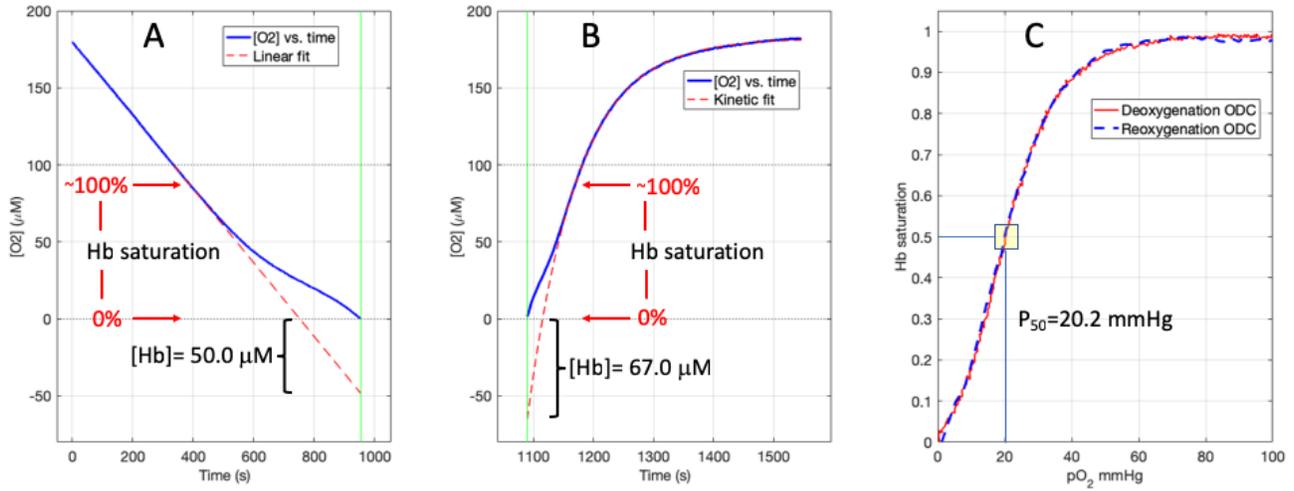

**Fig. S1. Graphic derivation of Hb non-equilibrium ODC.** Same deoxygenation/reoxygenation trace as in **Fig. 1b**. **a.** Deoxygenation phase with a fit of the linear part of the trace solely due to $O_2$ consumption by mitochondria. Linear extrapolation of the fit to the time in which all the $O_2$ is used up (vertical green line) reveals the sigmoidal curve of Hb deoxygenation as the difference between the observed curve (blue line) and the linear fit (dashed red line). **b.** Reoxygenation phase with a non-linear fit of the part of the curve that is solely due to the back diffusion of $O_2$ in the cell. Extrapolation of the fit to the time in which the cell is reopened (vertical green line) reveals the sigmoidal curve of Hb reoxygenation as the difference between the observed curve (blue line) and the fit (dashed red line). **c.** Deoxygenation and reoxygenation non-equilibrium ODC's derived as point-by-point difference between the $O_2$ traces and the extrapolated fits, as shown in panels **a** and **b**.

Since the reoxygenation of Hb occurs more rapidly (~65 seconds, panel B) than its deoxygenation (~560 seconds, panel A), the reoxygenation ODC (dashed blue line, panel C) is based on fewer experimental points, and is thus inherently less accurate than the deoxygenation ODC (continuous red line, panel C). Accordingly, the amount of Hb estimated from the reoxygenation phase is less accurate than that estimated from the deoxygenation phase. These shortcomings of the graphic fit are completely avoided with the global fit of both deoxygenation and reoxygenation phases with a single Hb concentration and a single set of rate constants, using kinetic models of $O_2$ binding to Hb as described in the Results section.

As the experiment was conducted at the atmospheric $P_{CO2}$ of 0.3 mmHg, the estimated $P_{50}$ of 20.2 mmHg using the graphic method is left shifted with respect to the standard values of 26 mmHg at a $P_{CO2}$ of 40 mmHg [1]. A correction for the difference in $P_{CO2}$ at pH of 7.4 based on the equation [2]:

$$P_{50,\Delta CO2} = P_{50} + 1.273 * 0.1 * (\Delta P_{CO2}) + 1.083 * 0.0001 * (\Delta P_{CO2})^2$$

would yield a $P_{50}$ value of 24.3 mmHg. Since Hb devoid of allosteric heterotropic factors was used in this experiment, based on the equation [2]:

$$P_{50,\Delta_{DPG}} = P_{50} + 795.63 * [\Delta_{DPG}] - 19660.89 * [\Delta_{DPG}]^2$$

the loss of 2,3DPG binding (estimated in a difference in erythrocyte concentration of ~3.2 mM [3]) would justify an additional correction of the $P_{50}$ value, with a right shift of ~2.3 mmHg.

**Supplementary Information References**